\documentclass[11pt]{article}
\usepackage{euscript}
\usepackage{amssymb}
\usepackage{amsfonts}
\usepackage{amsbsy}
\usepackage{amsmath}
\usepackage{epsfig}
\usepackage{amsthm}
\usepackage{amscd}
\usepackage{amstext}
\usepackage{verbatim}

\usepackage[pdftex]{hyperref}

\def\ben{\begin{equation}}
\def\een{\end{equation}}
\def\half{{\textstyle{1\over2}}}

   \let\d=\delta \let\e=\varepsilon
   \let\k=\kappa
     
\let\s=\sigma

\let\w=\omega \let\G=\Gamma

\def\nn{\nonumber}

\def\be{\begin{equation}}
\def\ee{\end{equation}}
\def\beq{\begin{equation}}
\def\eeq{\end{equation}}
\def\ba{\begin{array}}
\def\ea{\end{array}}

\def\dalemb#1#2{{\vbox{\hrule height .#2pt
       \hbox{\vrule width.#2pt height#1pt \kern#1pt
               \vrule width.#2pt}
       \hrule height.#2pt}}}

\newcommand{\bea}{\begin{eqnarray}}
\newcommand{\eea}{\end{eqnarray}}

\def\R{{{\Bbb R}}}

\def\En{{\mathcal{G}}}

\def\ocal{{\mathcal{O}}}

\begin{document}

\begin{center}

{ \LARGE {\bf Locally critical umklapp scattering and holography}}

\vspace{1cm}

{\large Sean A. Hartnoll$^\flat$ and Diego M. Hofman$^\sharp$}

\vspace{0.7cm}

{\it $^\flat$ Department of Physics, Stanford University, \\
Stanford, CA 94305-4060, USA \\}

\vspace{0.3cm}

{\it $^\sharp$ Department of Physics, Harvard University,\\
Cambridge, MA 02138, USA \\}

\vspace{1.6cm}

\end{center}

\begin{abstract}

Efficient momentum relaxation through umklapp scattering, leading to a power law in temperature d.c. resistivity, requires a significant low energy spectral weight at finite momentum. One way to achieve this is via a Fermi surface structure, leading to the well-known relaxation rate $\Gamma \sim T^2$. We observe that local criticality, in which energies scale but momenta do not, provides a distinct route to efficient umklapp scattering. We show that umklapp scattering by an ionic lattice in a locally critical theory leads to $\Gamma \sim T^{2 \Delta_{k_L}}$. Here $\Delta_{k_L} \geq 0$ is the dimension of the (irrelevant or marginal) charge density operator $J^t(\w,k_L)$ in the locally critical theory, at the lattice momentum $k_L$. We illustrate this result with an explicit computation in locally critical theories described holographically via Einstein-Maxwell theory in Anti-de Sitter spacetime. We furthermore show that scattering by random impurities in these locally critical theories gives a universal $\G \sim \left( \log \, \frac{1}{T} \right)^{-1}$.

\end{abstract}

\pagebreak
\setcounter{page}{1}

\section{Context: the many faces of the d.c. resistivity}

The d.c. resistivity of a system with a net charge density is ultimately tied to the rate at which the charge-carrying excitations can lose their momentum. For instance, if the UV theory is relativistically invariant, allowing us to relate frames by boosting, then, in the presence of a net charge density and if the total momentum is conserved, a hydrodynamic argument shows that the conductivity diverges at the lowest frequencies $\w \to 0$ as \cite{Hartnoll:2007ih}
\be\label{eq:hydro}
\sigma(\w) = \frac{\rho^2}{\epsilon + P} \left( \frac{i}{\w} + \d(\w) \right) \,.
\ee
Here $\rho, \epsilon, P$ are the charge and energy densities and pressure, respectively.
Using quantum field theory, this result can also be obtained from Ward identities \cite{Hartnoll:2007ip, Herzog:2009xv}. The zero temperature limit of this expression is simplified using $\epsilon + P = \mu \rho$.

It is of course intuitively plausible that a net charge will accelerate indefinitely under an applied electric field. Formulae such as (\ref{eq:hydro}) indicate that the d.c. conductivity is a subtle observable. Despite being a low energy quantity, it is sensitive to UV data such as the charge density and translation invariance. Computations of finite d.c. conductivities at nonzero density necessarily include the effects of momentum non-conserving terms. The classic example is Fermi liquid theory in the presence of irrelevant couplings to a lattice, which leads to a universal resistivity $r \sim T^2$, as we will briefly review. Similarly, the remarkable robustness of the observed linear in (low) temperature resistivity across a range of chemically distinct unconventional materials (for an overview see \cite{Sachdev:2011cs}) may suggest that the key physics there is also universal, i.e. describable within the framework of effective field theory. In this letter we will present a new way, distinct from Fermi surface kinematics, in which the UV sensitivity of the d.c. resistivity can be subsumed into a critical effective field theory.

Any loss of momentum is a question of timescales. To invalidate the conclusion (\ref{eq:hydro}), and hence achieve a finite d.c. resistivity, momentum must be effectively lost on the experimental timescale. Two natural ways to achieve this are firstly if the charge carriers of interest are parametrically diluted by a bath of other degrees of freedom, and secondly if the charge carriers interact with parametrically heavier degrees of freedom. In both cases the charge carriers can dump their momentum into the other degrees of freedom and the momentum will not be returned to the charge carries within the experimental timescale. Any reliable computation of a d.c. conductivity must hinge on an approximation analogous to the two just described.

There has been some recent success realizing the former of these scenarios via the holographic correspondence. Firstly in `probe brane' setups, where the charge carriers are parametrically diluted by critical neutral degrees of freedom \cite{Karch:2007pd}. An important class of these models, where the probe brane is described by the Dirac-Born-Infeld action, have been shown to have a low temperature resistivity scaling as $r \sim T^{2/z}$, in 2+1 dimensions, with $z$ the dynamical critical exponent governing the critical neutral modes \cite{Hartnoll:2009ns}. A second set of holographic models that realize similar physics are the locally critical non-Femi liquids of \cite{Faulkner:2009wj}. Here, a parametrically small fraction of the charged degrees of freedom are fermions with non-Fermi liquid dispersion relations due to interactions with a bath of fractionalized charged degrees of freedom \cite{Faulkner:2010tq, Sachdev:2010um}. The contribution of the non-Fermi liquid excitations to the resistivity goes like $r \sim T^{2 \nu_{k_F}}$, where $\nu_{k_F}$ is related to the, UV sensitive, scaling dimension of the fermionic operator in the low energy locally critical theory of the fractionalized degrees of freedom \cite{Faulkner:2010zz}.

The second scenario, involving interaction with parametrically heavy degrees of freedom, has the advantage that such degrees of freedom always exist in actual materials: as quenched random impurities and/or as a lattice of ions.  It also does not depend on a large $N$ limit in an essential way. Many of the most interesting materials appear to be very pure, and partially for that reason we will focus on scattering off an ionic lattice in this letter. In the final section we will also present a result for random impurity scattering. A periodic lattice degrades momentum via umklapp scattering processes, in which the lattice momentum is conserved, but the true momentum is not. In the following section we review the quantum field theoretic treatment of umklapp scattering.

\section{Momentum relaxation rate due to umklapp scattering}

In this section we will arrive at an expression for the momentum relaxation rate due to umklapp scattering, and obtain the d.c. conductivity with this input.

Consider a translationally invariant field theory at finite density such that the only conserved vector quantity is the total momentum. We recalled above that this results in an infinite d.c. conductivity. Now, perturb this theory by an irrelevant operator (so the IR is still described by the original fixed point) that breaks translational invariance:
\be\label{hamil}
H= H_0 - g \,\mathcal O(k_L) \,,
\ee
where $k_L$ represents the typical lattice momentum scale breaking translation invariance.

With the Kubo formula in mind, define the following inner product in the space of operators at finite temperature T \cite{Forster}:
\be
C_{AB}(t) \equiv  \left(A | e^{- i L t} | B \right) = \left(A(t) | B \right) \equiv T  \int_0^{\frac{1}{T}} d\lambda \langle A(t)^\dag B(i \lambda) \rangle \,.
\ee
Here $L$ is the Liouville operator $L = [H, \cdot ]$.
The usefulness of this definition becomes manifest if we Laplace transform this expression:
\be\label{ctab}
\widetilde C_{AB}(\omega) = \left(A\left| \frac{i}{\omega - L} \right| B \right) = \frac{T}{i \omega} \left[G_{AB}^R(\omega) - G_{AB}^R(i 0)\right] \,,
\ee
where $G^R$ is the retarded Green's function. Obtaining this result requires similar manipulations to those involved in proving the fluctuation-dissipation theorem \cite{Forster}. In this section we follow the conventions of \cite{Forster} for Green's functions.
Therefore, the d.c. conductivity is given by the Kubo formula as:
\be\label{eq:s}
\sigma = \lim_{\w \to 0} \frac{\text{Im} \, G_{\vec J \vec J}^R(\w)}{\w} = \frac{1}{T} \lim_{\w \to 0} \widetilde C_{\vec J \vec J} (\omega) \,.
\ee

We would like to have a perturbative expression that captures the leading contribution coming from the leading irrelevant correction introduced in (\ref{hamil}). The appropriate way to do this is given by the memory matrix formalism \cite{Forster}. A crucial part of this formalism is the inclusion of all conserved operators that overlap with $\vec J$ in matrix conductivities. In our case this is only the momentum $\vec P$. Intuitively, the conductivity diverges because the current operator has some overlap with the momentum operator, which is conserved. Once this is considered, we can write
\be\label{eq:hs}
\hat \sigma (\omega) = \hat \chi \cdot \left( \hat M(\omega) - i \omega \hat \chi \right)^{-1} \cdot \hat \chi \,,
\ee
where hatted quantities are two dimensional matrices with indices either $\vec J$ or $\vec P$. We have defined the static susceptibilities
\be
\chi_{AB} = \frac{1}{T} C_{AB}(0) \,,
\ee
and the `memory matrix' is found to be \cite{Forster}
\be\label{eq:mab}
M_{AB}(\omega) = \frac{1}{T} \left( \dot A \left| \mathcal Q \frac{i}{\omega - \mathcal Q L \mathcal Q} \mathcal Q \right| \dot B\right) \,,
\ee
where the operator $\mathcal Q$ projects onto the space of operators perpendicular to $\vec J$ and $\vec P$. To show that the $\chi_{AB}$ as defined are in fact the static susceptibilities, take the inverse Laplace transform of (\ref{ctab}) and use the sum rule for the Green's function. At this point, (\ref{eq:mab}) is an exact expression for the memory matrix $\hat M$ appearing in (\ref{eq:hs}).

We will shortly argue that at low frequencies, in the setup described, it is only the momentum component $M_{\vec P \vec P}$ that determines the conductivities. In that case we obtain
\be\label{eq:jj}
\sigma_{\vec J \vec J} =   \lim_{\w \to 0}  \frac{ \chi_{\vec J\vec P}^2}{M_{\vec P \vec P}(\w)} \equiv \frac{\chi_{\vec J \vec P}^2}{\chi_{\vec P \vec P}} \, \frac{1}{\Gamma} \,,
\ee
where
\be\label{eq:gam}
\Gamma =  \lim_{\w \to 0} \frac{M_{\vec P \vec P}(\w)}{\chi_{\vec P \vec P}} \,,
\ee
is the momentum relaxation rate, as we can see from the $\vec P\vec P$ component of $\hat \sigma$. The expression (\ref{eq:jj}) has the form of the Drude-type formula (\ref{eq:hydro}) in a relativistic theory, as Ward identities imply $\chi_{\vec J \vec P} = \rho$ and $\chi_{\vec P \vec P} = \e + P$ \cite{Herzog:2009xv}.

Let us now give an explicit expression for $M_{\vec P \vec P}$ and show that the other components of $\hat M$ are subleading. The definition of $\vec P$ implies
\be\label{eq:dp}
\dot{\vec P} = i [H, \vec P] = g \, \vec k_L \, \mathcal O(k_L) \,.
\ee
This is consistent with the fact that $\vec P$ is conserved before perturbing.
On the other hand, $\dot{\vec J}$ is generically an order zero in $g$ operator evaluated at zero momentum, as the current is not conserved. Free theories are exceptions to this genericity statement, as is, to leading order at low energies, 2+1 dimensional Fermi liquid theory with a convex Fermi surface \cite{FS2}. Therefore, to leading order in $g$
\be\label{eq:mpp}
M_{\vec P \vec P} =  \frac{1}{T} \lim_{\w \to 0} \left( \dot{\vec P} \left| \frac{i}{\omega - L} \right| \dot{\vec P} \right)  
= \frac{g^2 k_L^2 }{T}  \lim_{\w \to 0} \left. \widetilde C_{\mathcal O \mathcal O}(k_L) \right|_{g=0} = g^2 k_L^2  \lim_{\w \to 0} \left. \frac{\textrm{Im}\, G_{\mathcal O\mathcal O}^R(\omega, k_L)}{\w} \right|_{g=0}  \,.
\ee
Notice that to lowest order we can perform all calculations involving $L$ and $\mathcal Q$ to zeroth order in $g$, as $\dot{\vec P}$ has already supplied an overall factor of $g^2$ from (\ref{eq:dp}). Therefore, momentum conservation, and in particular the fact that the inner product between states with and without momentum vanishes, implies $\mathcal Q  = 1$. The power of the memory function method is precisely that it identifies the matrix $\hat M$ as the quantity that can reliably be treated perturbatively at late times. $\hat M$ captures the effects of all conserved quantities, while non-conserved quantities play no essential role at late times. As we argued above $M_{\vec J \vec J}$ must be order zero in $g$. What about the mixed term? Naively one would say it is order $g$. This is not the case, however. Momentum conservation at leading order makes this term vanish and the result is $M_{\vec J \vec P} \sim g^2$.
Putting these observations together we obtain
\be\label{eq:mmm}
 \left(\hat M\right)^{-1}_{\vec P \vec P} \sim M_{\vec P \vec P}^{-1} \sim g^{-2} \,, \quad\quad \left(\hat M\right)^{-1}_{\vec J\vec J}   \sim M_{\vec J \vec J}^{-1} \sim 1\,, \quad\quad \left(\hat M\right)^{-1}_{\vec J \vec P} \sim - \frac{ M_{\vec J \vec P}}{M_{\vec J \vec J}M_{\vec P \vec P}} \sim 1 \,.
\ee
This demonstrates the claimed property that $M_{\vec P \vec P}$ dominates the low frequency conductivities (\ref{eq:hs}).
We expect a perturbative analysis (of the memory matrix) to be valid at low energies provided that the coupling $g$ to the lattice is irrelevant, as we have assumed.

A case of direct interest to transport in strongly correlated metals is umklapp scattering by an ionic lattice. In that case the operator $\ocal = J^t$ and the lattice appears as a spatially dependent chemical potential. Putting the above formulae together, the momentum relaxation rate is
\be\label{eq:relax}
\G =  \frac{g^2 k_L^2}{\chi_{\vec P \vec P}} \lim_{\w \to 0} \left. \frac{\text{Im} \,  G^R_{J^t J^t}(\w,k_L)}{\w} \right|_{g=0} \,.
\ee

\section{Critical umklapp with and without Fermi surfaces}

The upshot of the previous section is that the momentum relaxation rate, and hence the conductivity, due to perturbative umklapp scattering by an ionic lattice is given through the density-density spectral function, $\text{Im} \, G^R_{J^t J^t}(\w,k)$, with $\w \to 0$ and $k = k_L$, the lattice momentum. In order for this quantity to be captured by a critical effective field theory, with, say, a resistivity that is a power law in the temperature, it is necessary that low energy excitations exist at $k = k_L$. If no excitations are supported at the lattice momentum, for instance if all excitations have a dispersion towards the origin in momentum space $\w \sim k^{z}$, then the resistivity will be due to Boltzmann excitated states only and will be exponentially small at low temperatures. 

Systems with a Fermi surface can naturally admit critical umklapp scattering in two senses, as we now review. The first is if the umklapp momentum connects two points on the Fermi surface. Then all charge carriers involved in the umklapp scattering are critical, despite the momentum transfer. This process is mediated by the density operator at finite momentum. In 2+1 dimensions
\be\label{eq:jtfl}
J^t(\w,k) \equiv J^t(p) = \int d^3q \, \psi^\dagger_\sigma (q) \psi_{\sigma} (p+q) \,.
\ee
In Fermi liquid theory this operator has dimension $\Delta = -1$ under scaling towards the Fermi surface (i.e. $\w, k_\perp$ scale but $k_\parallel$ does not). It is therefore relevant. It will induce an RG flow that will fold the Fermi surface and gap out the two points connected by the lattice vector. This conclusion can be averted either by tuning the gap to be zero or by non-Fermi liquid physics that renders the operator (\ref{eq:jtfl}) irrelevant.
An interesting example of the first possibility is given by the `hot spots' on a Fermi surface coupled to a critical spin density wave. Fermions at the hot spots contribute a strong power law conductivity, but can easily be short-circuited by the remainder of the `cold' fermions \cite{Rice, Rosch}. A careful renormalization group treatment of the system \cite{Metlitski:2010vm} suggests, however, that the critical umklapp scattering at the hot spots can be communicated to the rest of the Fermi surface \cite{Hartnoll:2011ic}. In 1+1 dimensional systems, such hot spots constitute the entire Fermi `surface' and one might expect that e.g. a half-filled Luttinger liquid could exhibit a critical resistivity due to umklapp scattering, in cases where the umklapp coupling is irrelevant \cite{Giamarchi}. This expectation is thwarted by additional conservation laws in 1+1 dimensions that require two distinct umklapp process in order to relax the current, and these two process cannot both be critical \cite{ra1,ra2,ra3}.

The second way in which Fermi surface kinematics enable critical umklapp scattering is through coupling the lattice to a more complicated operator than the charge density. The irrelevant quartic interaction of Fermi liquid theory can easily be generalized to violate momentum conservation by some lattice momentum $k_L$. In 2+1 dimensions
\be
S_4 = g \int dt \left( \prod_{i=1}^4 d^2k_i \right) \psi^\dagger_\sigma (k_1) \psi^\dagger_{\sigma'} (k_2) \psi_\sigma (k_3) \psi_{\sigma'} (k_4)
\d(k_1 + k_2 - k_3 - k_4 - k_L) \,. 
\ee
The RG flow is towards the Fermi surface. In particular, the delta function does not scale generically \cite{Polchinski:1992ed, Shankar:1993pf}. At low energy the $k_i$ are constrained to take values on the Fermi surface. Consider for instance the case of a circular Fermi surface. It is easy to see pictorially that, for all $k_L$ with magnitude between $0$ and $2 k_F$, it is possible to satisfy both the Fermi surface constraint and the delta function constraint in the interaction with scatterings involving fermions at all points on the Fermi surface. The entire Fermi surface will have a critical resistivity. This is a well known fact, and the corresponding momentum relaxation is easy to compute in the framework of the previous section. The operator
\be
\ocal(\w,k) \equiv \ocal(p) = \int \left( \prod_{i=1}^4 d^3p_i \right) \psi^\dagger_\sigma (p_1) \psi^\dagger_{\sigma'} (p_2) \psi_\sigma (p_3) \psi_{\sigma'} (p_4) \d^{(3)}(p_1 + p_2 - p_3 - p_4 - p)  \,,
\ee
has scaling dimension $\Delta = 1$ under scaling towards the Fermi surface. The Fourier transform with respect to energy, $\ocal(t,k)$, therefore has dimension $\Delta=2$ and the frequency space Green's function $G^R_{\ocal\ocal}(\w,k)$ has dimension $\Delta = 4-1 = 3$. This last step uses the fact that delta functions of generic momenta do not scale under RG towards the Fermi surface \cite{Polchinski:1992ed, Shankar:1993pf}. The imaginary part of the Green's function must be odd under $\w \to -\w$ and therefore we can anticipate the momentum relaxation rate from (\ref{eq:gam}) and (\ref{eq:mpp}) will have the (low) temperature dependence
\be
\G \sim \lim_{\w \to 0} \frac{\text{Im}\, G^R_{\ocal\ocal}(\w,k_L)}{\w} \sim T^2 \,.
\ee
This is the well-known Fermi liquid theory result.

The previous paragraph is a little too fast. For a convex Fermi surface in 2+1 dimensions, the current $\vec J$ remains a conserved quantity in the presence of the quartic interactions of Fermi liquid theory, working to leading order at low frequencies. In these cases our scalings (\ref{eq:mmm}) will not hold true, as we assumed that the current $\vec J$ was degraded prior to the consideration of umklapp effects. A proper treatment considers $\vec J$ and $\vec P$ simultaneously
via the entire memory matrix \cite{FS2,FS1} and again recovers a resistivity $r \sim T^2$.

The above considerations can straightforwardly be generalized to cases where the excitations of the Fermi surface do not have Fermi liquid dispersion relations. All that potentially changes is the dimension of the fermionic operators under scaling towards the Fermi surface. Thus, for example, one could add umklapp scattering to the models of  \cite{Faulkner:2009wj}.

Without a Fermi surface-like structure (including e.g. Fermi points), one is left with scalings towards the origin $\w \sim k^z$. An exceptional case, however, is the limit $z \to \infty$. In this limit, time scales but space does not. In such a locally critical theory, {\it all} momenta become independently critical at low energies.
It is immediately clear that umklapp scattering off an ionic lattice will lead to critical resistivities in such a theory. The charge density operator $J^t(\w,k)$ will have a scaling dimension $\Delta_{k}$ under the critical scaling. The UV quantity $k_L$ will then determine the IR scaling dimension $\Delta_{k_L}$ of the modes that control the loss of momentum. 
With this difference, that $k_L$ appears in the operator dimension, the logic then proceeds very similarly to the Fermi liquid case. In particular, the momentum conservation delta function again does not scale, leading to $G^R_{J^t J^t}(\w,k)$ having dimension $2 \Delta_{k}  + 1$. Therefore, the temperature dependence of the momentum relaxation rate (\ref{eq:relax}) is
\be\label{eq:localumklapp}
\G \sim \lim_{\w \to 0} \frac{\text{Im} \,  G^R_{J^t J^t}(\w,k_L)}{\w} \, \sim \, T^{2 \Delta_{k_L} } \,.
\ee
If we require the operator $J^t(\w,k)$ to be marginal or irrelevant in the IR theory -- and if this is not the case then we have not reached the true IR and our perturbation theory is suspect -- then $\Delta_{k_L} \geq 0$. As for the Fermi liquid,
a marginal operator leads to a constant, $T^0$, momentum relaxation rate.

Locally critically theories also dovetail in an interesting way with Fermi surfaces, as one can efficiently scatter fermionic excitations with locally critical bosons. This fact is behind the non-Fermi liquid spectral functions and resistivities of \cite{Faulkner:2009wj, Sachdev:2010um, Faulkner:2010zz}. In this work we are exploiting a different consequence of local criticality: the efficiency of umklapp scattering in such a theory, independently of the presence of Fermi surfaces.

In the following section we will compute the resistivity due to umklapp scattering in a concrete holographic model that exhibits local criticality in the IR.

\section{Holographic model}

The holographic correspondence geometrizes the renormalization group flow \cite{Heemskerk:2010hk, Faulkner:2010jy}. In particular, the IR field theoretical physics is described by the far interior of the dual spacetime. In the absence of explicit charged matter in the bulk \cite{Hartnoll:2011fn}, it is a robust feature \cite{Sen:2005wa} that at zero temperature and at finite charge density, a fully regular solution to the bulk equations of motion will have an $AdS_2 \times \R^2$ IR geometry. It was emphasized by \cite{Faulkner:2009wj} that the isometries of this IR spacetime -- time is part of the $AdS_2$ factor and scales while space does not -- entailed an emergent local criticality. In fact, the scaling in time is part of a larger emergent $SL(2,\R)$ symmetry of $AdS_2$ that strongly constrains low energy Green's functions, as we will see shortly. Here, as in the remainder, we have specialized to 2+1 dimensional field theories.

The simplest model that illustrates the physics of interest is Einstein-Maxwell theory in asymptotically Anti-de Sitter spacetime:
\be\label{eq:act}
S = \int d^4x \sqrt{-g} \left( \frac{1}{2 \k^2} \left(R + \frac{6}{L^2} \right) - \frac{1}{4 e^2} F_{\mu \nu}F^{\mu \nu}  \right) \,.
\ee
We are interested in the universal low energy physics of models with holographic duals of the form (\ref{eq:act}). In particular, we need to compute
retarded Green's functions at low temperature and frequencies $\w, T \ll \mu$. The momentum however need not be small.
It is well established that holographically the dissipative low frequency physics is captured by the near-horizon geometry, while low temperatures means that the horizon will be near-extremal (see e.g. \cite{Faulkner:2011tm}). Therefore, we can focus on the following solution to the theory, which describes a black hole in $AdS_2 \times \R^2$:
\be\label{eq:ads2}
ds^2 = \frac{L^2}{6} \left( - \frac{f(r) dt^2}{r^2} + \frac{dr^2}{f(r) r^2} + dx^2 + dy^2 \right) \,, \qquad A = \frac{1}{\sqrt{6}} \frac{e L}{\k} \left(\frac{1}{r} - \frac{1}{r_+} \right) dt \,.
\ee
Here the metric function
\be
f(r) = 1 - \frac{r^2}{r_+^2} \,.
\ee
This background is the near-horizon geometry that captures the finite temperature, locally quantum critical physics.

To compute the retarded Green's function of $J^t$ in this background at finite energy $\w$ and momentum $k$ we
must perturb the time component of the Maxwell potential: $\d A_t$. Due to the finite energy and momentum, this perturbation
couples to other modes. Taking the momentum to be in the $x$ direction, without loss of generality, we find that we need to consider the perturbations
\be
\{\d g_{xx}, \, \d g_{yy}, \, \d g_{tt}, \, \d g_{xt}, \, \d A_t, \, \d A_x\} \,.
\ee
All the perturbations have the form of a function of $r$ times $e^{- i \w t + i k x}$. The Einstein-Maxwell equations for the perturbations are easy to write down but rapidly become unwieldy. Fortunately, it has been found that a clever choice of gauge invariant variables can reduce these equations to two decoupled second order equations \cite{Kodama:2003kk, Edalati:2010pn}. For our background, we must first introduce
\be
\Phi = \d A_t' - \sqrt{\frac{3}{2}} \frac{\d g_{tt}}{f} \,, \qquad \Psi = \d g_{yy} \,,
\ee
and then define
\be
\Phi_\pm = \Psi + \frac{r^2}{\sqrt{6} k^2} \left(1 \pm \sqrt{1 + 2 k^2} \right) \Phi \,.
\ee
These variables are now found to satisfy
\be
\Phi_\pm'' + \frac{f'}{f} \Phi'_\pm + \left( \frac{\w^2}{f^2} - \frac{1 + k^2 \pm \sqrt{1 + 2 k^2}}{r^2 f} \right) \Phi_\pm = 0 \,.
\ee
These equations can be solved in terms of hypergeometric functions. Imposing, as usual \cite{Son:2002sd, Hartnoll:2009sz}, infalling boundary conditions at the horizon, we obtain
\bea
\Phi_\pm & = & (2 r_+)^{\nu_\pm} \G\left(a_\pm \right) \G\left( 1 + \nu_\pm \right) f(r)^{- i r_+ \w/2} r^{\frac{1}{2} - \nu_\pm}
{}_2F_1\left(\frac{a_\pm}{2}, \frac{a_\pm+1}{2}, 1 - \nu_\pm ;\frac{r^2}{r_+^2} \right) \nn \\
& & \qquad \qquad - \left(\nu_\pm \leftrightarrow - \nu_\pm \right) \,, \label{eq:ads2sol}
\eea
with $a_\pm = \half - i r_+ \w - \nu_\pm$. We also defined the exponents \cite{Edalati:2010pn}
\be\label{eq:nupm}
\nu_\pm = \frac{1}{2} \sqrt{5 + 4 k^2 \pm 4 \sqrt{1 + 2 k^2}} \,.
\ee
This formula looks a little different to the formula in \cite{Edalati:2010pn} because we have normalized the spatial coordinates $x,y$ differently in (\ref{eq:ads2}).

To match the near-horizon solution (\ref{eq:ads2sol}) onto the solution away from the horizon, one expands the solution near the boundary $r \sim 0$ of the $AdS_2 \times \R^2$ region
\be\label{eq:matching}
\Phi_\pm \propto r^{\frac{1}{2}} \left(r^{- \nu_\pm} + \En_\pm(\w) \, r^{\nu_\pm} \right) \,.
\ee
The locally quantum critical Green's functions are found to be
\be\label{eq:gt}
\En_\pm(\w) = - (\pi T)^{2 \nu_\pm} \frac{\G\left(1 - \nu_\pm\right)\G\left(\frac{1}{2} - \frac{i \w}{2 \pi T} + \nu_\pm
\right)}{\G\left(1 + \nu_\pm\right)\G\left(\frac{1}{2} - \frac{i \w}{2 \pi T} - \nu_\pm\right)} \,.
\ee
We have given this result in terms of the temperature of the black hole
\be
T = \frac{1}{2 \pi r_+} \,.
\ee
The expression (\ref{eq:gt}) is in fact determined up to overall normalization, which we have not been overly careful about, by the scaling dimensions $\nu_\pm + \half$ of the operators and the $SL(2,\R)$ symmetry of the black hole in $AdS_2$ \cite{Faulkner:2011tm, de Alfaro:1976je, Spradlin:1999bn}.
For the momentum relaxation rate due to umklapp scattering, we will be interested in the $\w \to 0$ expansion of the imaginary part of the Green's function. This gives
\be\label{eq:smallw}
\text{Im} \, \En_\pm(\w) = \w \left(\pi T \right)^{2 \nu_\pm - 1} \frac{\pi}{2} \frac{\G(1-\nu_\pm) \G(\half + \nu_\pm)}{\G(1+ \nu_\pm) \G(\half - \nu_\pm)} \tan \pi \nu_\pm + \cdots \,.
\ee
Note that this expression is regular for all positive $\nu_\pm$. The poles and zeros in the tangent and gamma functions cancel each other out. The opposite limit, $T \to 0$, recovers the zero temperature Green's functions of \cite{Edalati:2010pn}. The finite temperature Green's functions have recently been studied numerically in \cite{Davison:2011uk}.

Away from the near horizon region, the fact that we are taking $\w,T \ll \mu$ allows us to simply put $\w = T = 0$ into the perturbation equations. The decoupling of the equations is trickier in the full geometry, the spacetime is no longer a direct product as in (\ref{eq:ads2}), but has been achieved in \cite{Edalati:2010pn}. We do not need to know the solution to these equations. A well-established matching procedure, see e.g. \cite{Faulkner:2009wj,Edalati:2010pn}, in the overlap regime where the solutions behave as (\ref{eq:matching}),
gives the full low frequency Green's function as
\be
G_\pm(\w) = \frac{A + B \En_\pm(\w)}{C + D \En_\pm(\w)} \,,
\ee
where $A,B,C,D$ are real constants that are generically nonzero and independent of $\w$ and $T$. In particular, this implies that at the lowest frequencies
\be\label{eq:im}
\text{Im} \, G_\pm(\w) \, \propto \, \text{Im} \, \En_\pm(\w) \,.
\ee
From this result we can obtain the desired density-density Green's function. The near-boundary behaviour of $\d A_t$ can be extracted from the near boundary behaviour of the gauge invariant variables \cite{Edalati:2010pn}. The density-density Green's function is found to be a linear combination of the $G_\pm$ Green's functions with coefficients independent of frequency and momentum. At low temperature and frequencies, the $G_-$ Green's function is more singular than the $G_+$ Green's function, because $\nu_- < \nu_+$, and so gives the dominant contribution. One therefore finds from (\ref{eq:smallw}), (\ref{eq:im}) and (\ref{eq:relax}) that
\be
\Gamma \sim \lim_{\w \to 0} \frac{\text{Im} \,  G^R_{J^t J^t}(\w)}{\w} \, \sim \, T^{2 \nu_- - 1} \,.
\ee
This result is consistent with the general expression (\ref{eq:localumklapp}) and the fact that the dimension of the frequency-space operator is $\Delta_{k} = \nu_- - \half$. The following figure shows the power of the resulting d.c. electrical resistivity, $r \sim T^{2 \nu_- - 1}$, as a function of the lattice momentum $k_L$.
\begin{figure}[h]
\begin{center}
\includegraphics[height=130pt]{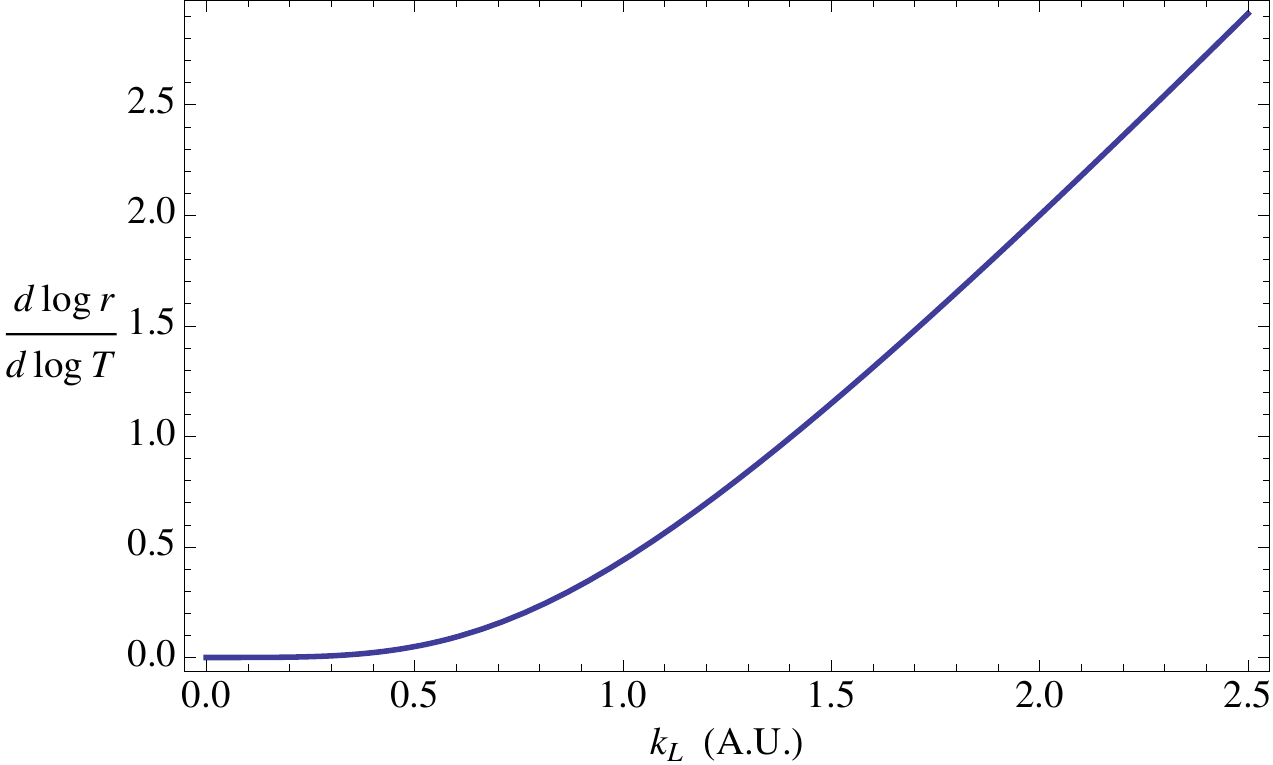}\caption{Temperature power law of the d.c. resistivity as a function of the lattice momentum $k_L$ in models with an Einstein-Maxwell gravity dual.
The physical normalization of the momentum depends on the UV completion of the locally critical theory.}
\end{center}
\end{figure}

The momenta $k$ appearing in the exponents (\ref{eq:nupm}) are dependent on the normalization of spatial coordinates. To obtain invariantly defined (UV dimensionless) momenta using only data from the near horizon region of the spacetime, we can consider the ratio
\be
\bar k^2 = \frac{k^2}{\rho_\text{fr.}} = \frac{e \k}{L} \sqrt{6} \,  k^2 \,.
\ee
Here $\rho_\text{fr.}$ is the electric flux emanating from the extremal $AdS_2$ horizon. This has the field theoretic interpretation \cite{Hartnoll:2011fn, Hartnoll:2011pp, Iqbal:2011bf} as the portion of the total charge density $\rho$ carried by `fractionalized' degrees of freedom, in the sense of \cite{Huijse:2011hp}.

\section{Discussion}

The primary observation of this letter is that local quantum criticality provides a new route, different to Fermi surface kinematics, to obtain critical umklapp scattering. It has been widely noted that locally critical theories, with $z=\infty$, are unlikely to describe stable phases of matter down to arbitrarily low temperatures due to their extensive ground state entropy and divergent density of states, e.g. \cite{Hartnoll:2009sz, Jensen:2011su}.
Nonetheless, locally critical theories may provide a useful description of certain systems over a range of intermediate energy scales, e.g. \cite{Sachdev:2010um, McGreevy:2010zz,  Iqbal:2011in, Iqbal:2011ae}. In particular, holography is likely to provide cases where this intermediate energy range can be made parametrically wide in a large $N$ limit.

At present we do not know how to obtain critical umklapp scattering from critical theories without Fermi surfaces and with $z <  \infty$. It may be impossible. One intriguing fact suggests that if it is possible to do so, a rather universal resistivity may be obtained. It has been observed that in multiple
different holographic settings with finite $z$ IR criticality, the optical conductivity $\s(\w) \sim \w^2$ \cite{arXiv:0908.3677, arXiv:0911.3586, arXiv:1008.2828}, using the technique introduced by \cite{Gubser:2008wz}. Dimensional analysis of the Kubo formula and charge conservation equation can then be used to show that the charge density operator $J^t(t,x)$ is marginal in all of these critical theories, with dimension $\Delta = 1+ 2/z$. This statement is also true in locally critical theories when the operator carries zero momentum \cite{Edalati:2009bi}. However, we have seen in (\ref{eq:nupm}) that a nonzero momentum increases the weight in this case, making the operator irrelevant. At finite $z$ the momentum is dimensionful and so cannot change the scaling dimension of the operator. Because the charge density operator at finite momentum controls umklapp scattering by an ionic lattice, it is conceivable that a universally marginal charge density could lead to a universal critical d.c. conductivity in these cases.

The memory function formalism extends to cases with critical degrees of freedom residing on the lattice sites themselves.
This situation has recently been considered holographically using a lattice of fermionic D-brane defects \cite{Sachdev:2010um, Jensen:2011su, Kachru:2009xf, Kachru:2010dk,  Sachdev:2010uj}. The works \cite{Jensen:2011su, Kachru:2010dk} computed a critical d.c. conductivity by coarse graining the lattice and then using the framework of \cite{Faulkner:2009wj, Faulkner:2010tq, Faulkner:2010zz} -- the conductivity is finite because the charged degrees of freedom considered constitute a parameterically small portion of the total degrees of freedom. To determine the effect of umklapp scattering in these models, one integrates out the degrees of freedom at the lattice sites to obtain lattice momentum couplings of the form (\ref{hamil}). The coupling `constants' $g$ so generated will be functions of $\w/T$ with a scaling determined by the dimension of the lattice operator.
This is a physically distinct mechanism for momentum relaxation compared to \cite{Jensen:2011su,Kachru:2010dk}.
%, and leads to different temperature power laws. For instance, a marginal impurity operator `semi-holographically' coupled to free fermions generates the umklapp operator (\ref{eq:jtfl}), but now with a coupling $g \sim T^2 \, F(\w/T)$ and hence leading to a resistivity $r_\text{umk} \sim T^2$.

In this work we have not touched upon the computation of optical conductivities. The optical and d.c. conductivities are deeply interconnected but behave in opposite ways. As we have seen, many low energy degrees of freedom can lead to a large resistivity, and hence small d.c. conductivity. On the other hand, the optical conductivity is essentially the spectral density for charged degrees of freedom and is therefore large when the d.c. conductivity is small. Critical optical conductivities due to umklapp scattering were recently found in an RG treatment of the quantum critical spin density wave transition in two dimensions \cite{Hartnoll:2011ic}. Various interesting optical conductivities have also been obtained in holographic models without umklapp, e.g. \cite{Hartnoll:2009ns,Faulkner:2010zz, Hartnoll:2011dm} as well as the $\w^2$ results mentioned above. In considering these results, one can distinguish conceptually between frequency dependence due to the Drude-like broadening of the delta function in (\ref{eq:hydro}), increasing at small frequencies, and that due to the depletion of spectral weight at nonzero frequencies by the Drude peak, decreasing at small frequencies. It is of interest to extend our results to a systematic consideration of possible critical optical conductivities due to umklapp scattering.

Finally, we recall that scattering off random impurities is also naturally treated using the memory function method. A formula for the scattering rate in that case was obtained in \cite{Hartnoll:2007ih,Hartnoll:2008hs}. The formula is essentially just an integral of our expression (\ref{eq:localumklapp}) over momenta, which we might think of as averaging over lattice separations. For the case of charged impurities
\be
\G_\text{imp} \sim \lim_{\w \to 0} \int \frac{d^2k}{(2\pi)^2} k^2 \frac{\text{Im} \,  G^R_{J^t J^t}(\w,k)}{\w} \sim \int dk \, k^3  \, T^{2 \Delta_{k} }\,.
\ee
In the low temperature limit, this integral is dominated by the small momentum contribution. Using the concrete expression (\ref{eq:nupm}),
the momentum is found to have a natural scale $k^4_\star \sim \left(\log \, \frac{1}{T} \right)^{-1}$, which is small. A scaling argument then gives the momentum relaxation rate
\be
\G_\text{imp} \sim \frac{1}{\log \frac{1}{T}} \,.
\ee
Unlike the umklapp scattering we have focussed on, this relaxation rate is completely universal in the sense that it does not depend on the UV completion of the locally critical IR theory. The power of the logarithm that appeared is sensitive to the fact that in the holographic model $\Delta_k \sim k^4$ at small $k$. A different power in the small momentum expansion would have led to a different power of the logarithm.

\section*{Acknowledgements}

It is a pleasure to acknowledge discussions with Chris Herzog, John McGreevy, Subir Sachdev, Eva Silverstein and especially Max Metlitski while this work was in progress. We also benefited greatly from the stimulating environment of the KITP in Santa Barbara, during the workshop on Holographic Duality and Condensed Matter Physics. This research was supported in part by the National Science Foundation under Grant No. NSF PHY05-51164, by DOE grant DE-FG02-91ER40654, and the Center for the Fundamental Laws of Nature at Harvard University. The work of S.A.H. is partially supported by a Sloan research fellowship.

\end{document}